\documentclass[runningheads]{llncs}
\usepackage[T1]{fontenc}
\usepackage{graphicx}
\usepackage{multirow}
\usepackage{amsmath,amssymb,amsfonts}
\usepackage{booktabs}
\usepackage{multirow}
\usepackage{graphicx}
\usepackage{hyperref}
\usepackage{caption}
\usepackage{subcaption}
\begin{document}
\title{SDLFormer: A Sparse and Dense Locality-enhanced Transformer for Accelerated MR Image Reconstruction}
%
%
\author{Rahul G.S.\inst{1,2} \and
Sriprabha Ramnarayanan\inst{1,2} \and
Mohammad Al Fahim \inst{1,2} \and Keerthi Ram \inst{2} \and {Preejith S.P} \inst{2} \and
Mohanasankar Sivaprakasam \inst{2} }
\authorrunning{Rahul et al.}
\institute{Indian Institute of Technology Madras (IITM), India \and
2 Healthcare Technology Innovation Centre (HTIC), IITM, India
\email{\{rahul.g.s,sriprabha.r\}@htic.iitm.ac.in}}
\maketitle              
\begin{abstract}
Transformers have emerged as viable alternatives to convolutional neural networks owing to their ability to learn non-local region relationships in the spatial domain. The self-attention mechanism of the transformer enables transformers to capture long-range dependencies in the images, which might be desirable for accelerated MRI image reconstruction as the effect of undersampling is non-local in the image domain. Despite its computational efficiency, the window-based transformers suffer from restricted receptive fields as the dependencies are limited to within the scope of the image windows.
We propose a window-based transformer network that integrates dilated attention mechanism and convolution for accelerated MRI image reconstruction. The proposed network consists of dilated  and dense neighborhood attention transformers to enhance the distant neighborhood pixel relationship and introduce depth-wise convolutions within the transformer module to learn low-level translation invariant features for accelerated MRI image reconstruction. The proposed model is trained in a self-supervised manner. We perform extensive experiments for multi-coil MRI acceleration for coronal PD, coronal PDFS and axial T2 contrasts with 4x and 5x under-sampling in self-supervised learning based on k-space splitting. We compare our method against other reconstruction architectures and the parallel domain self-supervised learning baseline. Results show that the proposed model exhibits improvement margins of (i) $\sim$ 1.40 dB in PSNR and $\sim$ 0.028 in SSIM on average over other architectures (ii) $\sim$ 1.44 dB in PSNR and $\sim$ 0.029 in SSIM over parallel domain self-supervised learning. The code is available at 
\url{https://github.com/rahul-gs-16/sdlformer.git}

\end{abstract}
\section{Introduction}

\begin{figure}[t!]
    \centering
    \includegraphics[trim={0.07cm 0.25cm 0.33cm 0.25cm},clip,scale=0.55]{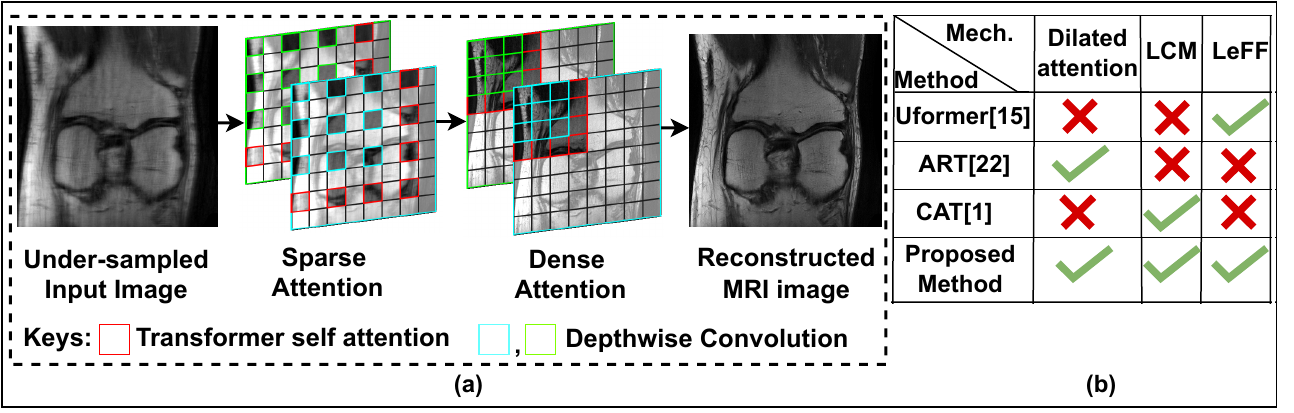}
    
    \caption{Description of the proposed model. In the sparse attention block pixels are given in dilated manner to the input of the transformer. In dense attention block the input is given in a continuous manner. Cyan and green denote different channels.
    }
    \vspace{-0.5cm}
    \label{fig:concept_diag}
\end{figure}
Vision transformers have emerged as a competitive alternative to convolutional blocks in various image reconstruction tasks \cite{liang2021swinir},\cite{zamir2022restormer},\cite{wang2022uformer},\cite{zhou2023dsformer}. They offer a flexible mechanism to capture relationships between regions from distant neighborhoods \cite{vaswani2017attention}, helping in relating patterns useful for image restoration. MRI acceleration can specifically benefit from this, as the imaging process involves sampling k-space trajectories, which impacts the image domain representation in a non-local manner. 

In this work, we consider the problem of MR image reconstruction using the window-based self-attention mechanism of vision transformers. 
Window-based transformers such as SwinMR \cite{huang2022swin} have been used for MRI reconstruction, but
windowing trades-off restricted receptive field for computational complexity, which we propose to alleviate by 
designing a variant of the transformer module. Further, we complement the global information operations of transformers with  convolutions, imparting fine-grained local feature modeling, valuable in MRI reconstruction. (Fig \ref{fig:concept_diag} (a)).

\textbf{Related works:} To increase the range over which attention is computed without increasing the computation cost, the Attention Retractable Transformer (ART) \cite{zhang2022accurate} uses a sparse attention block (SAB) where the input to the transformer is the windowed image regions processed in a dilated manner. 
 On the other hand, the Cross Aggregation Transformer (CAT) \cite{chen2022cross} introduces the Locality Complementary Module (LCM) in the self-attention stage of the transformer to provide local context, 
 but processes the input using dense neighboring windows. 
 An alternative method of inducing locality is Locality enhanced Feed-Forward (LeFF) \cite{wang2022uformer}, which uses depth-wise convolutions in the transformer's feed-forward layer. 
 While recursive dilated CNNs \cite{sun2018compressed} have been applied in MRI reconstruction, an intersection of dilated or sparse self-attention mechanism with local context remains unexplored for MRI reconstruction. 
 Figure ~\ref{fig:concept_diag}b tabulates the key factors and motivations for our proposed method.

Our proposed model is a window-based self-attention transformer that incorporates sparse and dense attention blocks with convolutions. Designed to capture long-range pixel interactions and local contextual features, the proposed model is trained in a data-driven self-supervised manner \cite{yaman2020self}, and demonstrated for 4x and 5x accelerations in multi-coil MRI. 

We summarise our contributions as follows. 
\textbf{1)} We propose SDLFormer, a computationally efficient and performant transformer-based network hybridized with CNNs for accelerated multi-coil MRI reconstruction.
\textbf{2)} Our proposed transformer block is designed to capture long-range dependencies via sparse attention on dilated windows in the input image domain and dense attention over neighborhood windows. The sparse and dense attention blocks are augmented with depth-wise convolutions to learn local contextual features with self-attention.
\textbf{3)} We extensively evaluate the proposed network on self-supervised multi-coil MRI reconstruction using multi-coil knee datasets for three contrasts: Proton Density (PD), Proton Density with Fat Suppression (PDFS), and Axial T2. We have achieved an improvement of  $\sim$ 0.6 dB in PSNR and  $\sim$ 0.018 in SSIM over the best-performing model, the SwinMR transformer. We perform an ablative study to understand the contribution of each component of our proposed transformer network.


\section{Method}

In this section, the mathematical formulation of the MRI under-sampling process, the overall architecture pipeline, and the Locality Enhanced Transformer (LET) block are described.

\textbf{Problem formulation:} Let $x \in \mathbb{C}^{N_x \times N_y}$ represent 2-D MRI image with height $N_y$ and width $N_x$. 
The forward model of the k-space undersampling process with $N_c$ coils is given by,

\begin{equation}\label{eqn:usampling}
    y_i=M \odot \mathcal{F}\left(S_i \odot x\right) ; \quad i=1, \ldots, N_c 
\end{equation}
where $M$ is a 2-D under-sampling mask , $\odot$ represents Hadamard product, and $\mathcal{F}$ is 2-D Fourier transform respectively. $S_i$ represents the sensitivity map that encodes the coil's spatial sensitivity and is normalized such that $\sum_{i=1}^{N} S_i^{*} S_i = I_n$

Our goal is to reconstruct image $x$ from $y$ which is formulated as an optimization problem for supervised learning given by,
\begin{equation}
    \label{eqn:optimization}
    \underset{\theta}{\operatorname{argmin}} \quad ||x - h_{\theta}(y)||_{2}^2 + 
     \lambda||M \odot \mathcal{F}(x) - y||_{2}^2
\end{equation}
where $x_u = \mathcal{F}^{-1}(y)$ is the undersampled image obtained by zero filling the missing k-space values, and $h_\theta$ is  the image reconstruction network. 

\textbf{Self-supervised learning:} Following \cite{yaman2020self}, we randomly partition $y$ into two disjoint sets $y_1$ and $y_2$  as follows $y_1 = M_1 \odot y, y_2 = M_2 \odot y$, 
where $M_1$ and $M_2$ are the two disjoint masks used to partition the k-space $y$. The loss $L$ function is defined as,  
\begin{equation}
    L(y_1,y_2) =  || M_2 \odot \mathcal{F}(h_{\theta}(y_1)) - y_2||_{1} 
\end{equation}
This self-supervised approach eliminates the need for fully sampled data, which requires extensive amounts of measurements for multi-coil acquisition. 



\begin{figure*}[t!]
    \centering
    \includegraphics[trim={0.15cm 3.0cm 0.15cm 0.25cm},clip,width=\linewidth]{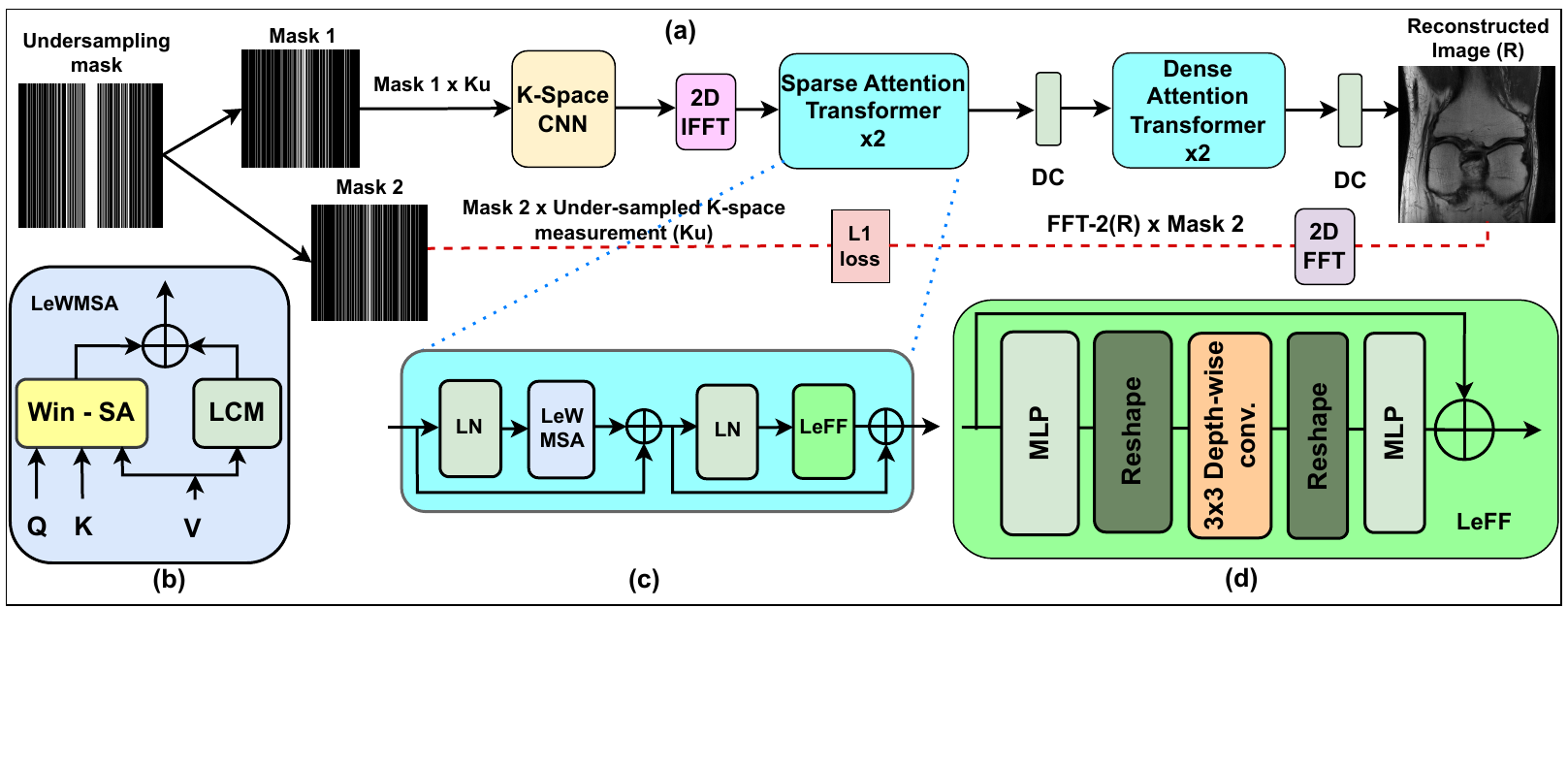}
    
    \caption{Proposed Overall Pipeline. The pipeline consists of k-space CNN, Sparse attention block, dense attention block, and convolution layer in series. LeWMSA, LeFF, and LCM represent Locally enhanced Window Multi-head Self Attention, Locally enhanced Feed Forward, and Locality Complementary Module respectively. The under-sampling mask is split into two and multiplied with the k-space measurement to obtain two subsets of measured k-space. One subset is used by the model for prediction and the other subset is used as reference. L1 loss is computed between the reference k-space subset and the FFT of the image predicted by the model.}
    \label{fig:pipeline}
\end{figure*}

\textbf{Architecture Details:} 
The overall pipeline of the proposed method is shown in Figure \ref{fig:pipeline} (a). The input to the pipeline is the under-sampled k-space data which is processed using a k-space CNN. The output of k-space is converted to the image domain. Initially, two Sparse Attention Transformer modules are present, followed by two Dense Attention Transformer modules. The LET block operates as the transformer module in the sparse and dense attention blocks. The sparse attention differs from the dense attention transformer by operating in a dilated manner. K-Space CNN is a 5-layer CNN with instance normalization, ReLU activation, and a residual connection from the input to the output to enable gradient flow. The pipeline of the architecture is shown in Figure \ref{fig:pipeline} (a). 

\textbf{Locality Enhanced Transformer (LET):} The internal architecture of the LET block is shown in Figure \ref{fig:pipeline} (c). This architecture tries to address two main challenges faced by a vanilla transformer. 1) The quadratic computation cost with respect to the number of tokens. 
2) The transformers show a limitation in capturing local dependencies \cite{wu2021cvt}, \cite{li2021localvit} which are essential for image restoration.


Computational complexity is reduced using a non-overlapping Locality enhanced Window-based Multi-head Self-Attention (LeW-MSA). Here the input feature map $X \in \mathbb{R}^{(H \times W \times C)}$ is split into non-overlapping windows with window size $M \times M$ and flattened to obtain features $X_{i} \in \mathbb{R}^{M^2 \times C}$ from each window i. The flattened features are projected into subspace using linear projection to obtain query (Q), key (K), and Value (V). 
Multi-headed self-attention is applied, to the flattened features in each window using the equation \ref{eqn:wmsa} a.
Inspired by CAT \cite{chen2022cross}, a Locality Complementary Module (LCM) is introduced in the transformer, which is a 3x3 depth-wise convolution module, used to extract local contextual features, as shown in Figure \ref{fig:pipeline} (b). Following Uformer \cite{wang2022uformer},\cite{yuan2021incorporating}, the ability of the transformer's feed-forward layer to capture local contextual features is improved by introducing a 3 x 3 depth-wise convolution in the feed-forward layer, with the necessary reshaping as shown in Figure \ref{fig:pipeline} (d). The architecture of the transformer block is shown in Figure \ref{fig:pipeline} (c).


\begin{subequations}\label{eqn:wmsa}
\begin{align}
Attention(Q,K,V) = Softmax(\frac{QK^{T}}{\sqrt d_{r}} + B)V + LCM(V), \\
X' = LeWMSA(X_{in}) \ + \ X_{in}, \\
X_{out} = LeFF(LN(X')) \ + \ X' 
\end{align}
\end{subequations}


Where $X^{’}$ and $X_{out}$ are the outputs of Window-based Multi-head self-attention and $LeFF$ blocks with skip connections respectively. $LN$ represents Layer Norm.  
\textbf{Dataset details:} Three protocols: coronal proton-density (PD), coronal fat-saturated PD (PDFS), and axial fat-saturated T2 from the dataset Multi-Coil Knee dataset \cite{hammernik2018learning}, were chosen. The data was acquired through a 15-channel multi-coil setting for 20 subjects. Each 3D volume has 40 slices of 640x368 resolution complex-valued data and their corresponding sensitivity maps. The center 19 slices were considered for our experiments. The dataset was partitioned into 10 volumes containing 190 slices each for training purposes, and 10 volumes with 190 slices each for validation.

\textbf{Implementation details:}
The models are trained using PyTorch v1.12 on a 24GB RTX 3090 GPU.  The Adam optimizer \cite{kingma2014adam} without weight decay is employed with $\beta_{1} = 0.9$, $\beta_{2} = 0.999$, and an initial learning rate of 1e-3, which undergo step-wise reduction using a learning rate scheduler with a step-size of 40 epochs and $\gamma$ of 0.1. 
The training is performed for 150 epochs using the L1 loss, and the instances with the best validation loss are saved for comparison. Performance evaluation is based on Peak Signal-Noise Ratio (PSNR) and Structural Similarity Index Measure (SSIM).  Since aliasing artifacts are structural, non-local processing in the image domain alone might be insufficient \cite{9607653}. To address this, k-space CNN is added at the beginning of the proposed method and all the other comparison methods as suggested by \cite{eo2018kiki}, \cite{zhou2020dudornet}, \cite{ryu2021k}. For faster convergence, the weights of K-space CNN in the proposed model are initialized with the weights obtained by training KIKI-net in a self-supervised manner. The weights of the transformer are initialized with the weights obtained from the Uformer \cite{wang2022uformer} model trained with natural images (SIDD dataset). The data consistency (DC) \cite{schlemper2017deep} in the network's output is ensured using the partition $y_1$. The model is also trained in parallel domain training methodology \cite{hu2021self} and evaluated.

\section{Results and Discussion}
Our results are organized as follows.
1. Comparison of the proposed model with other State of the Art MRI reconstruction models.
2. Ablative study of various components in the architecture

\subsection{Qualitative and Quantitative Comparison on
Multi coil knee MRI dataset.}

\begin{table}
\caption{Quantitative comparison of different methods on different datasets and acceleration factors. The best results for self-supervised learning \cite{yaman2020self} are highlighted in bold. The SL, PL, and SSL represent Supervised Learning, self-supervised learning in the Parallel domain training, and Self Supervised Learning with only k-space splitting respectively.}
\resizebox{\textwidth}{!}{%
\begin{tabular}{@{}ccccccc@{}}
\toprule
\multirow{2}{*}{Method} & \multicolumn{2}{c}{Coronal PD}                                                                                                & \multicolumn{2}{c}{Coronal PDFS}                                                                                              & \multicolumn{2}{c}{Axial T2}                                                                                                   \\ \cmidrule(l){2-3} \cmidrule(l){4-5} \cmidrule(l){6-7} 
                        & \begin{tabular}[c]{@{}c@{}}4x\\ PSNR / SSIM\end{tabular} & \begin{tabular}[c]{@{}c@{}}5x\\ PSNR / SSIM\end{tabular} & \begin{tabular}[c]{@{}c@{}}4x\\ PSNR / SSIM\end{tabular} & \begin{tabular}[c]{@{}c@{}}5x\\ PSNR / SSIM\end{tabular} & \begin{tabular}[c]{@{}c@{}}4x\\ PSNR / SSIM\end{tabular} & \begin{tabular}[c]{@{}c@{}}5x\\ PSNR / SSIM\end{tabular} \\ \midrule
ZF & 
    28.14 / 0.7838 & 
    25.99 / 0.7119 & 
    30.67 / 0.7848 & 
    28.84 / 0.7206 &
    31.35 / 0.8186 &
    30.38 / 0.7829 \\ 
Recurrent VarNet \cite{yiasemis2022recurrent} &
    29.49 / 0.8255 &
    25.65 / 0.7134 &
    30.40 / 0.7880 &
    28.42 / 0.7237 &
    32.44 / 0.8402 &
    31.26 / 0.7983 \\ 
VS-Net \cite{duan2019vs}  & 
    30.65 / 0.8431 &
    26.71 / 0.7369 & 
    30.59 / 0.7810 &
    28.76 / 0.7176 &
    32.17 / 0.8269 &
    30.82 / 0.7885 \\ 
KIKI-net \cite{eo2018kiki} & 
    31.80 / 0.8617 &
    27.42 / 0.7574 & 
    33.13 / 0.8286 & 
    30.36 / 0.7462 &
    34.12 / 0.8564 &
    32.65 / 0.8138 \\ 
ISTA-Net \cite{zhang2018ista} &
    31.94 / 0.8635 &
    27.72 / 0.7649 &
    32.43 / 0.8072 &
    29.54 / 0.7320 &
    33.73 / 0.8485 &
    31.82 / 0.8013 \\
ISTA-Net (PL) \cite{hu2021self} &
    32.10 / 0.8698 &
    27.66 / 0.7620 &
    32.38 / 0.8067 &
    29.56 / 0.7322 &
    33.73 / 0.8501 &
    31.59 / 0.7988 \\
U-Net \cite{zbontar2018fastmri}&
    32.90 / 0.8884 &
    27.90 / 0.7729 & 
    33.58 / 0.8318 &
    30.79 / 0.7561 &
    34.36 / 0.8596 &
    32.67 / 0.8152 \\ 
SwinMR \cite{huang2022swin} &
    33.22 / 0.8954 &
    \textbf{28.43} / 0.7853 &
    33.65 / 0.8303 &
    30.59 / 0.7508 &
    34.38 / 0.8596 &
    32.81 / 0.8157 \\ 
Proposed (SSL) &
    \textbf{33.77 / 0.9056} &
    28.42 / \textbf{0.8031} &
    \textbf{33.96 / 0.8359} &
    \textbf{30.90 / 0.7611} &
    \textbf{34.97 / 0.8651} &
    \textbf{32.97 / 0.8193} \\
Proposed (PL) &
    33.86 / 0.9065 &
    28.71 / 0.8085 &
    34.07 / 0.8358 &
    30.97 / 0.7604 &
    35.01 / 0.8656 &
    33.08 / 0.8186 \\
Proposed (SL) &
    36.16 / 0.9282 &
    30.97 / 0.8495 &
    35.09 / 0.8534 &
    32.15 / 0.7837 &
    36.01 / 0.8814 &
    34.24 / 0.8393 \\ \bottomrule
\end{tabular}%
}
\label{quant_mc}
\end{table}

The quantitative comparison of the proposed model with other models proposed for the Multi-Coil MRI image reconstruction is shown in Table {\ref{quant_mc}}. Our method outperforms other methods proposed for MRI reconstruction in PSNR and SSIM metrics on all three MRI sequences (coronal PD, coronal PDFS, and axial T2) in both 4x and 5x acceleration factors, except coronal PD with 5x acceleration in terms of PSNR. The proposed model outperforms the second-best model by 0.59 dB in PSNR and 0.0178 in SSIM, in the axial-T2 dataset for the acceleration factor of 4x and in coronal PD for the acceleration factor 5x respectively. It can be seen that in the parallel domain self-supervised training mode \cite{hu2021self}, the proposed model outperforms the ISTA-Net.

\begin{figure*}[t!]
    \centering
    \includegraphics[trim={0.25cm 10.5cm 0.55cm 0.25cm},clip,scale=0.36,width=\linewidth]
    {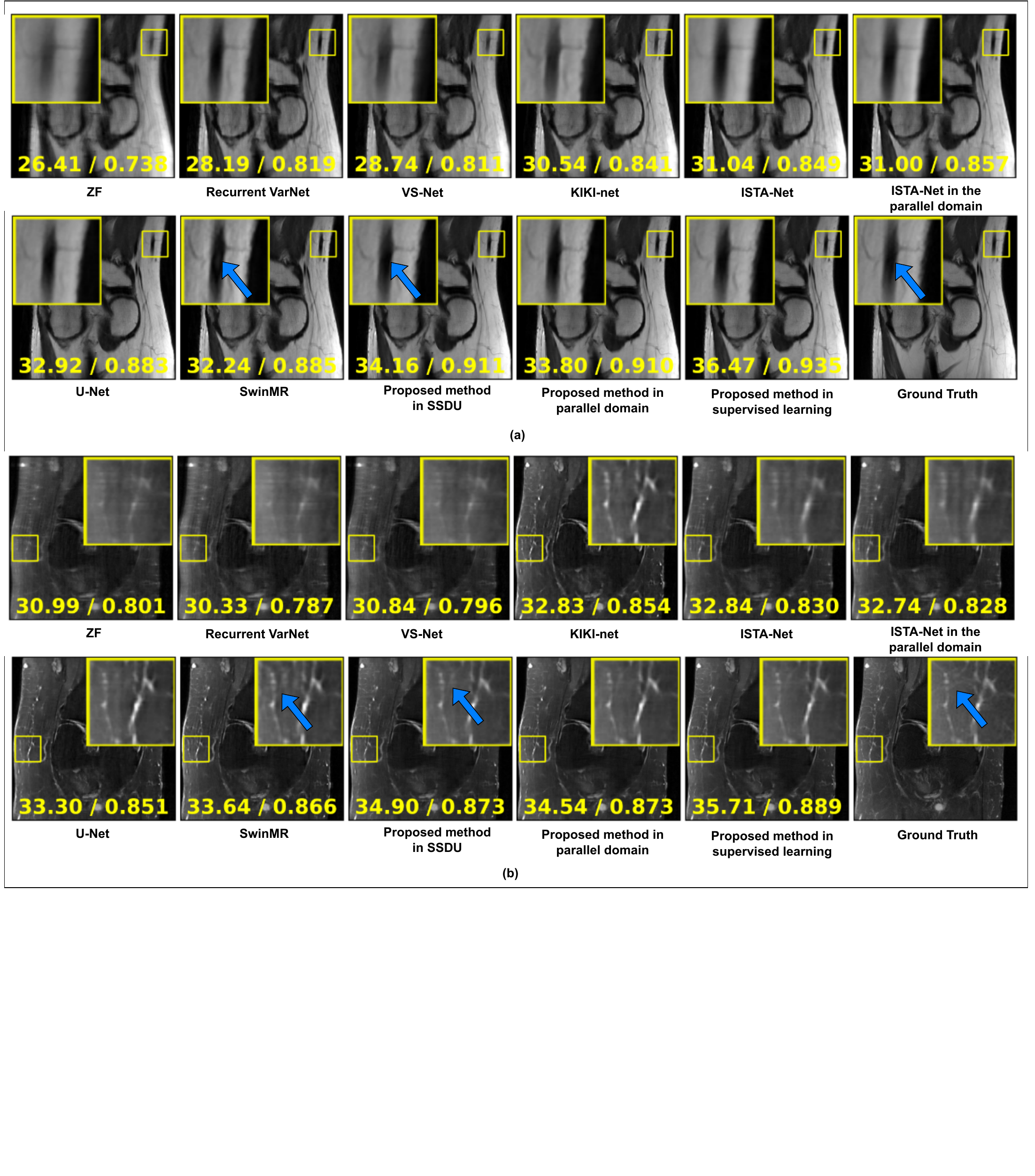}
    

    \caption{(a). Coronal PD 4x acceleration. Blue arrows highlight some artifacts produced by SwinMR which is not present in the ground truth and in the result of the proposed model. (b). Coronal PDFS 4x acceleration. Blue arrows highlight the sharp reconstruction produced by the proposed method.} 
    
    \label{fig:Coronal PD 4x acceleration}
\end{figure*}

The qualitative comparison of our model with other models proposed for multi-coil MRI reconstruction for coronal PD, axial T2 dataset for acceleration factors of 4x and 5x are shown in Fig \ref{fig:Coronal PD 4x acceleration} (a), \ref{fig:Coronal PD 4x acceleration}(b) respectively.



The reconstructions obtained through zero padding exhibit significant aliasing artifacts and lose important anatomical details. Although VS-Net, and Recurrent VarNet models are able to reduce the aliasing artifacts to some extent, the artifacts remain visible in the reconstructions. While the KIKI-net, U-Net, and ISTA-Net models are more effective in reducing the artifacts, they fall short in their ability to reconstruct the structures as accurately as transformer-based models. This can be attributed to their limited capability to capture long-range dependencies. Out of the transformer-based models, it can be seen that the proposed model reconstructs the image structures more accurately than the SwinMR transformer-based model. In Figure \ref{fig:Coronal PD 4x acceleration} (a), the SwinMR transformer-based model introduces some artifacts (region pointed to, by the blue arrow) in the image, which are not present in the ground truth. In Figure \ref{fig:Coronal PD 4x acceleration} (b), the proposed model recovers fine details better than the SwinMR model (as pointed to, by the blue arrow) when trained in self-supervised \cite{yaman2020self} \cite{hu2021self} or supervised techniques. From the results, it can be seen that increasing the receptive field by using dilated attention and complementing global information operations of transformers with fine-grained local contextual features from convolutions, positively impacts accelerated MRI image reconstruction.



\subsection{Ablation Study}

\begin{figure*}[h!]
    \centering
    \includegraphics[trim={0.25cm 0.27cm 4.2cm 0.25cm},clip,scale=0.25,width=\textwidth]{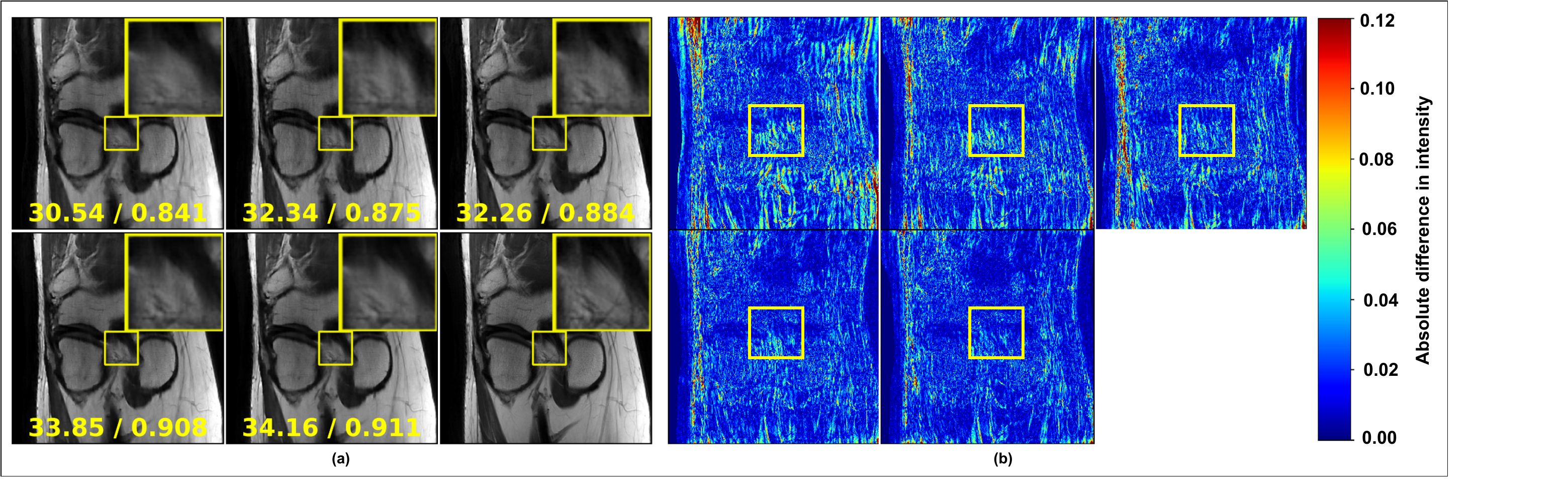}
    \caption{(a) Qualitative comparison of different models (b). Residual map to highlight the difference. Images from left to right correspond to the output of CNN, SAB, DAB, SAB and DAB without locality, Proposed method, and ground truth respectively.}
    \label{fig:ablation_qual}
\end{figure*}

The impact of each block in the network is analyzed in Table \ref{quant_blocks}. The results show that sparse attention blocks (SAB) significantly improve the model's performance. Dense attention block (DAB) individually performs better than SAB, but when they are combined together, they complement each other and provide the benefits of both attention methods. It can be seen that the locality-based enhancement improves the SSIM considerably, as it enables better capturing of local contextual features. This can be seen in Figure \ref{fig:ablation_qual}.
\begin{table}[t!]
\centering
\scriptsize
\caption{Comparison of different blocks in the model. The best results are highlighted in bold.}
\begin{tabular}{cc}
 
\hline
Architecture                                                     & PSNR / SSIM             \\ \hline
CNN                                                              & 31.80 / 0.8617          \\ \hline
SAB                                                              & 32.99 / 0.8863          \\ \hline
DAB                                                              & 33.10 / 0.8926          \\ \hline
SAB + DAB w/o locality                                           & 33.47 / 0.8974          \\ \hline
\textbf{SAB + DAB}                                               & \textbf{33.77 / 0.9056} \\ \hline
\end{tabular}
\label{quant_blocks}
\end{table}

\section{Conclusion}
Our work increases the receptive field of the transformer without increasing the computational complexity and complements global information with local contextual features by integrating convolutions in transformers. We have trained the proposed model in a self-supervised manner to remove the necessity of fully sampled data. We have evaluated our model for the reconstruction of the multi-coil knee MRI datasets in three different acquisition protocols and show that the proposed architecture outperforms other methods. We show that the integration of local contextual features obtained from convolution, with global information obtained using dilated self-attention improves the performance of the image reconstruction.

%
%
%
\bibliographystyle{splncs04}
\bibliography{references}
%




\end{document}